\documentclass[
nopreprint,
numerical 
]{jasatex}

\usepackage{graphicx}
\usepackage{dcolumn}
\usepackage{amsmath,amsfonts}
\usepackage{bm}

\begin{document}

\preprint{}

\title[Echo-Doppler with voice-induced vibration]{Physical understanding of an echo-Doppler test\\ with voice-induced vibration}

\author{Alessio D'Alessandro}
 \affiliation{Dipartimento di Fisica, Universit\`a di Genova \& INFN}
 \email{adales@ge.infn.it}

\author{Franco Rosso}
\affiliation{Istituto diagnostico polispecialistico Il Baluardo
s.p.a., Genova}
 \email{rosso.franco@libero.it}

\author{Massimo Calabrese}
\email{massimo.calabrese@hsanmartino.it}
\author{Giuseppe Minetti}
 \email{GiuseppeMinetti@libero.it}
\author{Alessandro Villa}
 \email{dr.willa@libero.it}
\affiliation{Centro di senologia - Policlinico Pammatone - Azienda Ospedaliera Universitaria S.~Martino, Genova}%

\date{\today}

\begin{abstract}
The physical understanding of a method of detecting
mammalian cancer via vocalization during a normal echo-Doppler test is provided.
The backscattered ultrasound frequency in the case of a vocal humming
resonating in the chest wall is computed: the overall effect is that the signal/noise ratio
could be easily improved at no cost.
Clinical results are to appear separately elsewhere.
\end{abstract}

\maketitle

\section{\label{sec:intro}Introduction}
Doppler ultrasonography is a non invasive, non-x ray diagnostic
procedure that can improve the accuracy of clinically and
mammographically detected abnormalities. It is widely used to
differentiate the different tissue components in breast and as a
problem-solving tool in the radiologist's armamentarium. A
transducer (the probe) sends a series of short ultrasound pulses
into the tissue and periodically pauses to listen for the returning
sounds: via the Doppler effect we can detect the direction, velocity
and turbulence of blood flow in a vascularization. The idea of
setting the target into forced vibration has also been exploited in
many different techniques (a general review of several is given in
Ref.~\cite{Yamakoshi}), for instance the mechanical properties of a
tissue under forced oscillation can be revealed. Occasionally
ultrasounds themselves (at low frequency) are used to force the
motion of our target.\cite{Fatemi}

The method we examine here is based on a Doppler ultrasound
measurement under a vocal-forced oscillation, with application to
breast scan. If the patient is asked to hum with voice, {\it i.~e.~}
vocalize,  during a normal echo-Doppler mammalian test, the sound
vibrations may set into motion the different connective tissues
inside breast. The Doppler ultrasound probe, which can detect
movement, may thus be able to provide a signal even stronger than
the one due to the natural blood flow.

In this paper we calculate the Doppler ultrasound signal reflected
back by a little ($\sim 5$~mm) spherical inclusion, when the latter
is entrained by the vibrations of a viscoelastic medium oscillating
at voice frequency. With a comparison with the typical Doppler
signal of the usual non-vibrating situation we show that a good
signal/noise ratio could be easily obtained.

\section{Evaluating minor contributions}
Let us consider a spherical inclusion (a lesion) in a viscoelastic
medium (the mammalian tissues). A sound wave propagating in an
elastic medium can of course carry on in translational motion any
object embedded in the medium. In the same time other effects may
appear, for instance the periodic pressure variations of sound can
make the object pulsate (shrink and swell) with the same frequency.
In the specific case of a Doppler ultrasound measurement in a tissue
already vibrating at the frequency of the human voice, another issue
may \emph{a priori} affect the measurement. The velocity of
propagation of ultrasounds in a moving (vibrating) tissue is indeed
different from that in a motionless medium. In principle ultrasounds
scattered back by the spherical inclusion (the target) and collected
back by the probe may have a significantly different frequency in
either situation.

Let us consider the effect of having a moving medium in the
unidimensional case, for sake of clearness. First suppose that both
the medium and the probe are at rest and blood in the lesion's
capillarization, by which the ultrasound is scattered back, is
moving with velocity $u$ (e.~g. toward the probe) with respect to
the motionless medium. If the emitted frequency is $\nu$, the
collected back frequency $\nu'$ is
\begin{equation}
\nu'=\nu \frac{1+u/c}{1-u/c} \sim \nu\left(1+\frac{2 u}{c}\right)
 \label{Dopplernomotion}
\end{equation}
where $c\sim1540$m/s is the sound speed in the human body.

 Now suppose that a sound wave is propagated in the medium, which now is vibrating
with velocity $v$: we can fairly assume that during the period in
which the ultrasound travels from and back to the probe $v$ is
constant. Actually for an object $5$ cm far, the ultrasound
traveling time is $\sim\frac{10 cm}{c}$, about $1/100$ of the period
of a normal $200$ Hz sound wave. We also suppose that the probe,
held by the operator, is still at rest and that blood in our target
is moving again with the same velocity $u$ toward the probe, just as
if our target would not have been entrained by the motion of the
surrounding medium. We have then in sequence: a source at rest (the
probe), the medium, moving with velocity $v$ toward the source, and
 the target, which is moving with velocity $u$ with respect to the probe.
The collected back frequency $\nu'$ is approximately
\begin{equation}
\nu' \sim \nu \frac{\left(1+\frac{u}{c+v}\right)}
{\left(1-\frac{u}{c-v}\right)}\sim \nu \left(1+
   (1+\frac{v^2}{c^2})\frac{2 u}{c}\right)
 \label{Dopplermotion}
\end{equation}
We have thus shown that corrections due to the simple motion of the
surrounding tissues are negligible (second order in $v/c$). In other
words, if the object is not entrained in translational motion by the
wave, the signals observed by the echo-Doppler device with
(\ref{Dopplermotion})  and without (\ref{Dopplernomotion}) the
medium vibration are nearly the same.\cite{note1} For the same
reason, if the object is entrained by the medium and moves with
velocity $u$ for whatever reason, the signal received by the Doppler
apparatus is always eq. (\ref{Dopplernomotion}) \emph{no matter what
the underlying tissue velocity is}. Voice has then no means of
influencing the Doppler measurement \emph{directly}, with a
modification of the ultrasound propagation speed as described above
or with a direct interaction with the device (whose operating
frequencies are 1000 times higher). The only effects we need to take
care of are the indirect ones: \emph{the voice sets  the inclusion
into motion in some way and the probe detects the velocity $u$ of
the latter} no matter how the former is scattered or propagated.

Hence let us focus on the possible causes of motion for the inclusion
and for each one let us calculate the respective velocity contribution $u$: the
relative signal $\nu'$ follows through (\ref{Dopplernomotion}).

First of all we will give a rough estimate of the pulsation of the
spherical lesion due to the periodic change in pressure\cite{note2} of the
surrounding medium with the sound wave. There are three important
consequences of the small dimensions of the lesion ($\sim$ 1cm)
w.r.t. the voice wavelength:
\begin{itemize}
\item The external pressure $p_{ext}$ of the surrounding medium is the
same for the whole inclusion;
\item The vocal sound wave may be regarded as plane;
\item The natural frequencies of resonance of the inclusion are much
higher than the frequency of the sound wave. This means that in the
equation of motion for the pulsating lesion the inertia term is
negligible w.r.t. the elastic term: as a consequence, internal and
external pressures are perfectly balanced ($p_{ext}=p_{int}:=p$)
at any moment.
\end{itemize}
Friction can be neglected in this analysis of the pulsation motion
because the absence of friction leads to the worst situation where
the pulsation velocity is the greatest.

We can thus write
\begin{eqnarray}
\frac{\partial p_{int}}{\partial \rho_{int}} \frac{d \rho_{int}}{d t} &=& \frac{d p_{ext}}{d t}=\frac{d p}{d t}\\
c^2_{int}  \frac{d \rho_{int}}{d t} = -c^2_{int} \frac{\rho_{int}}{V}  \frac{d V}{d t}&=& i \omega p\nonumber\\
-\frac{3 \rho_{int}}{R} c^2_{int} \frac{d R}{d t}&=&i \omega p \nonumber
\end{eqnarray}
where $c_{int}$ is the sound velocity inside the pulsating object and $\rho_{int}$, $R$, $V$
its density, radius and volume respectively. $\rho$ is the density of the surrounding medium and $\omega$ the sound wave
pulsation instead (a $e^{i\omega t}$ time dependence is always understood, here and in the rest of the paper).
For a plane sound wave we have $p = \rho v c$, where $v$ is the velocity perturbation propagating at speed $c$,
while $\frac{d R}{d t} = u_{puls}$ is the pulsation velocity of the lesion's interface, the one seen by the Doppler probe.
For the pulsation velocity we thus obtain (discarding the $i$ phase factor)
\begin{equation}
u_{puls} = \frac{\omega \rho v c R}{3 \rho_{int} c^2_{int}}
\end{equation}

Let us compare the latter quantity with the result of a pure translational motion. Suppose
for the moment that the lesion is carried with a velocity $u_{transl}$ which is not very different
form the bulk medium velocity $v$: this working hypothesis is reasonable since the
inclusion is not very different from the surrounding tissues.
In such a situation the medium ``does not sense'' the presence of the inclusion
and transfers to the object a moment per unit time
which is the very same necessary for moving an equal volume of medium, that
is $\rho V \frac{d v}{dt}$.
We obtain the ``dynamical Archimedes' law''
\begin{equation}
\rho_{int} V \frac{d u_{transl}}{dt} = \rho V \frac{d v}{dt}
\label{translrough}
\end{equation}
which allows us to write for a plane wave
\begin{equation}
u_{transl} = \frac{\rho}{\rho_{int}} v
\end{equation}
From this we compute the ratio
\begin{equation}
\frac{u_{puls}}{u_{transl}} = \frac{\omega c R}{3 c_{int}^2}\sim\frac{1}{100}
\end{equation}
for $\omega$ coincident with the typical human voice frequency.

From this we learn that the inclusion may be considered
as a rigid object, as the interface velocity contribution due to its
pulsation movement is much smaller than the translational velocity of the sphere.
\emph{The ultrasound frequency scattered back to the Doppler is
almost totally due to the contribution of the translational movement
of the object carried on by the sound wave.}

Thus we can finally proceed to give the exact formula as if the
oscillating object was a rigid sphere (i.e. not pulsing).

\section{The main translational contribution}
Our starting point is the relation \cite{Norris, Oestreicher} which
links the force $F$ of a rigid sphere oscillating with pulsation
$\omega$ in a medium:
\begin{equation}
F=\frac{4}{3}\pi R^3 \rho \omega^2 \left[ A_1 \frac{h_1(k R)}{k R} - 6 B_1 \frac{h_1(h R)}{h R} \right]
\label{exact}
\end{equation}
where $h_n$'s are the spherical Hankel function of the first kind,
$\rho$ is the density of the surrounding tissue (not that of the
oscillating object!) and $k$ and $h$ are the longitudinal and
transverse wavenumbers $k=\omega/c_L$, $h=\omega/c_T$ with
$c_L=\sqrt{(\lambda+2\mu)/\rho}$ and $c_T=\sqrt{\mu/\rho}$ (see
table \ref{mulambda} for typical values of $\mu$ and $\lambda$).
\begin{table}[!htbp]
\caption{\label{mulambda} Typical values \cite{Oestreicher} for $\mu=\mu_1+i\omega
\mu_2$ and $\lambda=\lambda_1+i\omega \lambda_2$}
\begin{ruledtabular}
\begin{tabular}{|c|c|}
\hline
$\rho$ & $1060$ kg/m\^{}3\\
\hline
$\mu_1$ & $2500$ N/m\^{}2\\
\hline
$\mu_2$ & $15$ N s / m\^{}2\\
\hline
$\lambda_1$ & $2.6\cdot10^{9}$ N/m\^{}2\\
\hline
$\lambda_2$ & $\sim 0$ (at acoustic frequencies)\\
\hline
\end{tabular}
\end{ruledtabular}
\end{table}

The (\ref{exact}) is obtained \cite{Norris,Oestreicher} from the
potential theory, by writing the general solution on a basis of
spherical Hankel functions. A general form of Hooke's law is the
constraint which relates the pressure with the displacement of the
lesion, so that both elasticity and viscosity are taken into account
and moreover a general slip condition at the interface is provided
\cite{Norris}. The coefficients $A_1$ and $B_1$ in case of no
slipping interface are \cite{Oestreicher}
\begin{eqnarray}
A_1 &=& - \frac{(3-3i h R - h^2 R^2) k^3 R^3 e^{-i k R} }{k^2 R^2 (1-i h R) +(2  - 2 i k R - k^2 R^2) h^2 R^2} u_0\nonumber \\
B_1 &=& \frac{(3-3 i k R - k^2 R^2) h^3 R^3 e^{-i h R}}{3 k^2 R^2 (1-i h a) + (2 - 2 i k R - k^2 R^2) h^2 R^2} u_0
\end{eqnarray}
where $u_0$ is the maximum amplitude of displacement of the sphere.
What really concerns us is the limit $k R \rightarrow 0 $, as the
sound wavelength is much longer than the radius $R$ of the lesion.
In this limit it is easy to show that (\ref{exact}) reduces to
\begin{equation}
F = 6 \pi \mu R u_0 \left(1-i h R -\frac{1}{9} h^2 R^2\right)
\label{exprstill}
\end{equation}
which is the exact expression for the problem \cite{Ilinskii} of an oscillating sphere \emph{in a
incompressible medium} (indeed the medium can be regarded as incompressible if the wavelength involved is large).
In the expression above we can recover the translational velocity, as $\mu u_0 \sim i \mu_2 \omega u_0 = \mu_2 u_{transl}$.

This is not the whole story yet, as $F$ is the force exerted on a
fluid with no other external forces by an oscillating object with a
given law of motion. The case of an object set in motion by a moving
(incompressible) fluid is different: first of all we
need\cite{Landau} to replace  $u_{transl}$ with the relative
velocity $u_{transl}-v$, where $v$ is the fluid velocity very far
from the lesion, or equivalently, the velocity the whole fluid would
have if there were no embedded objects in it, and then we have to
add the ``Archimedes' contribution'' $\rho V \frac{dv}{dt}$ we
discussed above. The final expression for the force acting on the
spherical lesion is then
\begin{equation}
F = -6 \pi \mu_2 R (u_{transl}-v) \left(1-i h R -\frac{1}{9} h^2 R^2\right) + \rho V \frac{d v}{dt}
\end{equation}
By Newton's law for the oscillating lesion $F=\frac{4}{3}\pi R^3
\rho_{int} \frac{d u_{transl}}{dt}$ and substituting
\begin{equation}
u_{transl} = \frac{ \frac{9 \mu_2}{2 R^2 \omega} \left(1-i h R -\frac{1}{9} h^2 R^2\right) + i \rho }{\frac{9 \mu_2}{2 R^2 \omega}\left(1-i h R -\frac{1}{9} h^2 R^2\right)+ i \rho_{int}} v
\label{traslformula}
\end{equation}
The above formula can be also obtained as the $k\rightarrow 0$, $\lambda+2\mu\rightarrow\infty$ limit
of equation (44) in Ref.~\cite{Norris2}, with the substitution $F_p \rightarrow \frac{4}{3}\pi R^3 \rho_{int} i w u_{transl}$
and $u_0 \rightarrow (i \omega)^{-1} v$. The same paper (equation (46)) gives also an expression for the displacement
due to a transverse wave perturbation
\begin{equation}
u_{transl}' = \frac{\frac{\mu_2}{2 R^2 \omega} }{\frac{\mu_2}{2 R^2 \omega} \left(1-i h R -\frac{1}{9} h^2 R^2\right)+ i \rho_{int}} v
\label{traslformula2}
\end{equation}
which is of the same order of magnitude as (\ref{traslformula}) at acoustic frequencies.

If we know the typical tissue vibrational velocity $v$, or
equivalently the vibrational amplitude, we can then find the
associated translational velocity $u_{transl}$ of the lesion and
through (\ref{Dopplernomotion}) the amount of signal detected by the
Doppler. We notice that in the (unrealistic) limit of large $\rho$
and $\rho_{int}$ we can recover the naive result (\ref{translrough})
from (\ref{traslformula}).

We can have a rough estimate for a typical value of the unperturbed
vibrational velocity $v$ of the mammalian tissue from
Ref.~\cite{Sundberg}: here accelerometers were put on several points
on the chest wall in order to detect the maximum amplitude of
vibration during singing. For a non-professional singer the maximum
displacement at sternum was $3-4~\mu m$, which at voice frequency
corresponds roughly to $v\sim 0.5$ cm/s; a precise measurement with
accelerometers placed near the Doppler probe during the test has not
been done yet. Nevertheless we notice that already $u_{transl}\sim
v\sim 0.5$ cm/s may be enough to be detected by typical Doppler
probes: for instance Esaote LA523 or LA435 are able to see such a
signal. In Ref.~\cite{Koski} we can find the minimum velocity
detected by the most common probes in the typical frequency range
(5-10 MHz) used for a mammalian echo Doppler test: our signal is
always above the threshold.

In conclusion the main contribution $u_{transl}$ arising from the
translational velocity seems to be enough to give the Doppler signal
an important improvement, which could cast in the range of
observability a signal which would not have been observed without
``vocal humming''.

\section{Conclusions}
We described the frequency signal for an echo-Doppler test in a
(mammalian) tissue vibrating at human voice frequency. We showed
that the main contribution improving the signal/noise ratio is that
of the translational motion of the different tissue components
caused by the voice vibration and that voice has no other means of
significantly influence the measurement (by e.g. changing the speed
of propagation of ultrasounds in the underlying tissue). An
important simplification is that the human voice frequency is low
compared to the traveling period of the signal so that the
vibrational velocity of the medium can be considered constant during
an acquisition period: the systematic effect of having ultrasounds
propagating in a moving medium is thus ruled out.

We calculated the amount of signal as a function of the vibrational
amplitude and compared it with a typical range for an echo-Doppler
device. These calculations seem to indicate that asking the patient
to vocalize during the echo-Doppler test could be a good method to
improve signal/noise ratio at no-cost. Of course an extensive
clinical testing should be done in order to confirm these
on-the-paper results.


\begin{thebibliography}{9}
\bibitem{note2}
For the sake of this analysis we shall consider an isotropic situation where pressure is direction-independent,
even if this is not the general situation in a solid.
\bibitem{note3}
A conjugation sign should be taken in order to be
consistent with the notation in Ref.~\cite{Oestreicher}. Here we keep the convention of
Ref.~\cite{Norris}.
\end{thebibliography}
\end{document}